\documentclass[smallextended]{svjour3}       % onecolumn (second format)
\smartqed  % flush right qed marks, e.g. at end of proof
\usepackage{graphicx}
\begin{document}

\title{The Faddeev-Yakubovsky symphony%\thanks{Grants or other notes
}


\author{Rimantas Lazauskas*         \and
        Jaume Carbonell %etc.
}


\institute{Rimantas Lazauskas \at
              IPHC, IN2P3-CNRS/Universit\'e Louis Pasteur BP 28, F-67037 Strasbourg Cedex 2, France\\
%              Tel.: +123-45-678910\\
%              Fax: +123-45-678910\\
              \email{Rimantas.Lazauskas@iphc.cnrs.fr}           %  \\
           \and
           Jaume Carbonell \at
              Institut de Physique Nucl\'eaire, Univ Paris-Sud, IN2P3-CNRS, 91406 Orsay Cedex, France
\email{carbonell@ipno.in2p3.fr}    }

\date{Received: date / Accepted: date}

\maketitle

\begin{abstract}
We  briefly  summarize the main steps leading to  the
Faddeev-Yakubovsky equations in configuration space for N=3, 4 and
5 interacting particles.
\end{abstract}

%\tableofcontents

\section{Introduction} \label{intro}

The Faddeev-Yakubovsky (FY) equations constitute a rigorous
mathematical formalism
 for solving the non relativistic  N-body problem of Quantum Mechanics.
They were motivated by the drawbacks of the fundamental
Schr\"{o}dinger equation in determining the physical solutions,
mainly those related with the scattering states, of a 3-body problem. FY
equations provide a framework enabling us to implement  additional constrains
to the solutions of the Schr\"{o}dinger equation, which guarantee their physical meaning.

FY equations should in principle  allow us to obtain the  exact
solutions of this fundamental equation for an arbitrary number of
interacting particles with the only input of the interparticle
potential. The term  "exact" must be here understood in the
numerical sense, i.e. allowing numerical methods that could
provide « accurate enough » solutions in a controlled converging
scheme. Only when such a possibility is ensured one would be able
to disentangle the consequences of a deficient interaction from
the consequences of an approximate solution, as was the main aim in
the historically first physical problem  at the origin of these
theoretical developments: the three-nucleon problem. In practice,
and more than 50 years after the publication of  Faddeev's seminal
work
\cite{Fad_JETP39_1960,Fad_61,Fad_65,Faddeev_Merkuriev_QSTh_1993,Yakubovsky_SJNP5_1967},
the number of interacting particles for which the FY equations
have been properly solved remains severely limited to N$\leq$5.

The formulation of FY equations can be considered as a starting
point of  a very active domain of theoretical physics, the so
called Few-Body Physics, where "Few" emphasizes that obtaining
exact solutions is a  very difficult task and one might never outreach
systems with small number of particles, when aiming the complete solution of the
Quantum Mechanical problem.  That is if we include the full complexity of the scattering solutions.

We will present in this contribution a brief summary of the main
steps in the formulation of these fundamental equations,
essentially in configuration space, starting from the pioneering
work of Faddeev up to the recent solutions of the 5-body problem
\cite{Rimas_5N_FBS_2018,Rimas_5N_PRC97_2018,LHC_PLB791_2019}. Our
main focus will be set  to  explain the generalisation procedure
when the number of interacting particles is increased,  describing in
parallel an  efficient and unified numerical method and  pointing out the natural limitations of this approach.

%%%%%%%%%%%%%%%%%%%%%%%%%%%%%%%%%%%%%%%%
\section{Shortcomings of Schr\"{o}dinger's equation in describing the 3-body scattering}\label{sec:1}

The problem faced by the Schr\"{o}dinger's equation  to describe
$N\ge3$ interacting particle has been realized since the very
beginning of Quantum Mechanics, can be summarized in the two following complementary aspects.

When formulated in  configuration space, the Schr\"{o}dinger
equation for N=3 particles interacting via pairwise interactions $V_{ij}$
\begin{equation}\label{Sch_3p}
(E-{\cal H}_0) \psi \!=\!  V  \psi
\quad
{\cal H}_0\!=\!-\sum_{i=1}^N {\hbar^2\over 2m_i}\Delta_{\vec{r}_i }
\quad
V\!=\!\sum_{i<j=1}^NV_{ij}=V_{12}+V_{23}+V_{23}
\end{equation}
is a second order partial derivative equation for a
function $\psi(\vec r_1,\vec r_2,\vec r_3)$  defined on
$\mathbf{R}^9$. Its solution is physical only once $\psi$ satisfy some appropriate boundary conditions (BC).

However, the scattering of one particle  on a 2-body bound state,
say   1+(23)  reaction,  may give rise to many asymptotic
channels, simultaneously coexisting,  which have different
behaviours. It turns to be impossible to implement the ensemble of
these complex asymptotic structures in a single function. One can
expect, for instance:
\begin{itemize}
\item In the elastic channel $1+(23)\to 1+(23)$ an asymptotic
behavior like
\begin{equation}
\Psi(\vec r_1,\vec r_2,\vec r_3) \sim \chi_e \left( {\vec r_2+\vec
r_3\over 2} - \vec r_1   \right)  \;  u_e(\vec r_2-\vec r_3),
\end{equation}
where  $u_e$ is the bound state wave function of
particle (23).\par Depending on the incoming energy, this channel
coexists with as many inelastic channels (eventually infinite) as
excited states of the (23)  subsystem are accessible at this
energy. All of them will have similar, but different, boundary
conditions like
\begin{equation}   \Psi(\vec r_1,\vec r_2,\vec r_3)
\sim \chi_i \left( {\vec r_2+\vec r_3\over 2} - r_1   \right)  \;
 u^*_i(\vec r_2-\vec r_3),
\end{equation}
where $u^*_i$ represents an excited state wave function.

\item In the $1+(23)\to 3+(12)$ rearrangement channel, one expects a solution having the form
\begin{equation}
   \Psi(\vec r_1,\vec r_2,\vec r_3) \sim \chi_r\left( {\vec r_1+\vec r_3\over 2} - \vec r_2   \right)  \;  u^*_r(\vec r_1-\vec
   r_2),
\end{equation} where $u^*_r$ is the bound state wave function of
particles (12).\par Corresponding expressions are expected also
for the $1+(23)\to 2+(13)$ rearrangement channels.

\item For the $1+(23)\to 1+2+3$ break-up channel one could guess
\begin{equation} \Psi (\vec r_1,\vec r_2,\vec r_3) \sim {e^{ik \rho} \over \sqrt \rho};  \qquad\quad \rho^2 = (\vec{r}_2-\vec{r}_3)^2+
  \left( {\vec r_2+\vec r_3\over 2} - \vec{r}_1 \right)^2.  \end{equation}
\end{itemize}
 This list still misses electromagnetic processes
like  radiative capture  as well as  relativistic effects like creation  or annihilation of  particles.

In  a consistent theory, it would be highly desirable {\it  (i)}
to deduce these different asymptotic behaviors from the equation
itself and  {\it (ii)} uncouple and impose them naturally as a
boundary condition in the corresponding regions of configuration
space. This is however not possible with eq. (\ref{Sch_3p}). It is
furthermore clear, that the Schr\"{o}dinger equation is not
flexible enough to account and treat  such a rich variety of physical situations which are present already for N=3.

\bigskip
On another hand, when formulated in momentum space, the
Schr\"{o}dinger equation, as well as its equivalent
Lippmann-Schwinger (LS) equation for the 3-body T-matrix,
\begin{equation}\label{LS_3p}
T= V+ VG_0(E)T ,   \qquad G_0(E) = \frac{1}{ E-H_0 + i\epsilon },
\end{equation}
does not need any kind of BC.
However the potential $V$ in momentum space is plagued with delta-like singularities on each parwise term $V_{ij}$, like for instance
\begin{equation}
   V_{23}(\vec k_1,\vec{p}_1,\vec{k}'_1,\vec{p}'_1)= v_{23} (\vec
   k_1,\vec{k}'_1)
   \delta(\vec p_1- \vec p'_1), \end{equation}
which propagate in all orders of the perturbative expansion of
(\ref{LS_3p})
\begin{eqnarray}
T&=& V_1 + V_2+V_3 \cr
  &+&  V_1G_0 V_1 + V_1 G_0 V_2 + V_1G_0 V_3 + V_2 G_0V_1 + V_2G_0V_2  +\dots \cr %  + V_2G_0V_3 + V_3G_0V_1 + V_3G_0V_2+V_3G_0V3
  &+&  V_1G_0 V_1 G_0V_1 + V_1G_0 V_1 G_0V_2 + + V_1G_0 V_1 G_0V_3 + \ldots   \label{T_pert}
 \end{eqnarray}
thus breaking the compactness of the integral kernel
\cite{Fad_65,Faddeev_Merkuriev_QSTh_1993}. Because of that, the
mathematical foundations of the theory, mainly the existence and
uniqueness of the solutions, are lost and we are left with an
inconsistent "theoretical no man's land".

%%%%%%%%%%
\section{Faddeev equations}

In order to edge these problems, L.D. Faddeev derived in 1960s a
set of equations, equivalent to (\ref{Sch_3p}), which constitutes
a rigorous mathematical framework for describing the variety of
physical situations involved in N=3. This seminal result was later
used by S.P. Merkuriev to derive the boundary conditions in
configuration space \cite{Merkuriev_71,Merkuriev_74} what allowed
the first solutions in the case of 3-nucleon problem with realistic interactions
\cite{GL_PRL29_1972,GLM_PRL33_1974,MGL_AP99_1976}.

A preliminary  step consists in isolating the intrinsic dynamics of the three-body Hamiltonian (\ref{Sch_3p}).
This is achieved by introducing the Jacobi coordinates
%\begin{eqnarray}
%\vec{x}_{\alpha}&=&\sqrt{2m_{\beta}m_{\gamma}\over m_0(m_{\beta}+m_{\gamma})}  (\vec{r}_{\beta}-\vec{r}_{\gamma})    \qquad \alpha=1,2,3\cr
%\vec{y}_{\alpha}&=& \sqrt{2m_{\alpha}(m_{\beta}+m_{\gamma})\over M  m_0}
%        \left[{m_{\beta}\vec{r}_{\beta}+m_{\gamma}\vec{r}_{\gamma}\over m_{\beta}+m_{\gamma}}-\vec{r}_{\alpha}\right] \label{Jacobi}
%\end{eqnarray}
\begin{eqnarray}
\vec{x}_{\alpha}&=&\sqrt{2\mu_{\beta,\gamma}\over m_0} \;
(\vec{r}_{\beta}-\vec{r}_{\gamma});    \qquad \alpha=1,2,3.\cr
\vec{y}_{\alpha}&=& \sqrt{2\mu_{\beta\gamma,\alpha} \over m_0} \;
\left( \vec{r}_{\alpha} - \vec{R}_{\beta\gamma} \right),
\label{Jacobi_3}
\end{eqnarray}
where  $(\alpha\beta\gamma)$ is a circular permutation of (123),
%${1\over \mu_{\beta,\gamma} } = {1 \over m_{\beta}} +{1\over m_{\gamma}}$
$\mu_{s,t}$ denotes the reduced mass of the system  formed by particle clusters $s$ and $t$,   $\vec{R}_{s}$ the center of mass of cluster $s$.
%$M=m_1+m_2+m_3$ the total mass of the system
An arbitrary mass  $m_0$ is introduced to fix the length scale. They are supplemented with the center of mass coordinate
\begin{equation}  M\vec R=m_1\vec r_1 + m_2\vec r_2 +m_3 \vec r_3;
    \qquad M=m_1+m_2+m_3. \end{equation}
In terms of them,
%the kinetic part of (\ref{Sch_3p}) takes the form
%\begin{equation}
% {\cal H}_0 = - {\hbar^2\over m_0} \left( \Delta_{\vec x_i} + \Delta_{\vec y_i} - {m_0\over2M} \Delta_{\vec R} \right)
%\end{equation}
% and
 the solution of (\ref{Sch_3p}), $\psi$  factorizes into an intrinsic part $\Psi$ and a plane wave for the center of mass motion
\begin{equation} \psi(\vec r_1,\vec r_2, \vec r_3)  % \equiv \psi(\vec x_i,\vec y_i,\vec R)
= \Psi(\vec x_i,\vec y_i) \; e^{i\vec{P}\cdot \vec{R}}.
\end{equation}

The intrinsic wave function $\Psi$, the only one in which we will be interested hereafter, is  a solution of
\begin{equation}\label{Sch_3p_I}
 [ E-H_0 ] \Psi(\vec{x}_i,\vec{y}_i) = V
\Psi(\vec{x}_i,\vec{y}_i),
\end{equation}
with (in ${\hbar^2/m_0}$ units)
\begin{equation}\label{H0_3p_I}
H_0 = - ( \Delta_{\vec x_i} + \Delta_{\vec y_i}).
\end{equation}
This trivial geometrical operation not only reduces the
dimensionality of the problem, from on $\mathbf{R}^9$ to
$\mathbf{R}^6$, but is the only way to properly disentangle the
intrinsic 3-particle energy the from its center of mass
excitations. Notice that there are 3 different sets of Jacobi
coordinates (\ref{Jacobi_3}) and that the form of  intrinsic free
Hamiltonian (\ref{H0_3p_I}) is independent of a particular choice.
They are related each other by orthogonal transformations
 \begin{eqnarray}
\vec{x}_{\alpha} &=& +c_{\alpha\beta} \vec{x}_{\beta} +
s_{\alpha\beta}  \vec{y}_{\beta},   \cr \vec{y}_{\alpha} &=&
-s_{\alpha\beta} \vec{x}_{\beta} + c_{\alpha\beta}
\vec{y}_{\beta}, \label{TCJ}
\end{eqnarray}
with
\begin{equation}\label{CHC}
c_{\alpha\beta} =
-\sqrt{\mu_{\alpha}\mu_{\beta}\over(1-\mu_{\beta})(1-\mu_{\alpha})},
\qquad s_{\alpha\beta} =  \epsilon_{\alpha\beta}
\sqrt{1-c_{\alpha\beta}^2},
%                 = \epsilon_{\alpha\beta} \sqrt{1-\mu_{\alpha}-\mu_{\beta}\over(1-\mu_{\beta})(1-\mu_{\alpha})}
%\quad \epsilon_{\alpha\beta}=(-1)^{\beta-\alpha}sig{(\beta-\alpha)}
\end{equation}
and $\epsilon_{\alpha\beta}=(-1)^{\beta-\alpha}sig{(\beta-\alpha)}$.

\bigskip
The  central idea of Faddeev  was to split  the wave function in a
sum of three terms, the so-called Faddeev components (FC),
\begin{equation}\label{Psi_3p}
 \Psi=\sum_{i<j=1}^N \Phi_{ij}= \Phi_{12}+ \Phi_{13}+\Phi_{23},
 \end{equation}
defined by
\begin{equation}\label{Phi_3p}
\Phi_{ij}= G_0(E) V_{ij} \Psi;   \quad\qquad G_0(E)= ( E-H_0
)^{-1},
 \end{equation}
 and so each of them associated with an interacting pair $V_{ij}$.

The ensemble of FC  obey the system of three coupled equations
\begin{equation} \label{FE_Phi}
 (E-H_0-V_{ij} ) \Phi_{ij}= V_{ij}  \sum_{kl\neq ij}\Phi_{kl},
\end{equation}
known as Faddeev equations for the 3-body problem, and provide a
solution of the corresponding Schr\"{o}dinger equation.
%\begin{eqnarray}
%   (E-H_0- V_{12}  ) \Phi_{12} = V_{12} (  \Phi_{13} + \Phi_{23} ) \cr
%   (E-H_0- V_{13}  ) \Phi_{13} = V_{13} (  \Phi_{23} + \Phi_{12} ) \cr
%   (E-H_0- V_{23}  ) \Phi_{23} = V_{23} (  \Phi_{12} + \Phi_{13} )  \label{FE_Phi}
%\end{eqnarray}
Its is indeed straightforward to see that if $\Phi_{ij}$ represent
a solution of eqs. (\ref{FE_Phi}), their sum (\ref{Psi_3p}) provides a
solution of eq. (\ref{Sch_3p}).

Notice that, although using relations (\ref{CHC}), each Faddeev
component can be expressed in any of the Jacobi coordinate sets
($x_{\alpha},y_{\alpha}$), it has  a natural expression in the one
on which the corresponding pair potential  has the simplest form,
$V_{ij}(x_k)$. This justifies the often used one-index notation
for the components $\Phi_i\equiv \Phi_{jk}$,  valid only for N=3.

The main advantage of the Faddeev equations is that  they provide
three functions, on which we can impose the proper boundary
conditions corresponding to the different channels. In the
asymptotic region of the configuration space ($x_k\to\infty$) the
three Faddeev equations, coupled by short range interactions
$V_{ij}(x_k)$, decouple from each other. This fact allows to
associate to each FC a well defined asymptotic structure which can
be used as boundary conditions for the system of differential
equations (\ref{FE_Phi}). For instance for the 1+(23)$\to$ 1+(23)
S-wave elastic scattering one can prove \cite{MGL_AP99_1976} that
\begin{equation} \begin{array}{l}
\Phi_1(x_1,y_1)   \cr  \Phi_2(x_2,y_3)  \cr
\Phi_3(x_3,y_3) \end{array}=
\begin{array}{lclrr}
u_1(x_1)   &\left[\sin{q_1y}\right. &+& \tan(\delta) \cos{q_1y_1}
&\left.\right]/(x_1y_1),  \cr
  &      0,                          &  &  &              \cr
&      0,                          &  &   &              \cr
\end{array}
\end{equation}
whereas when all  rearrangements channels are open one must impose
\begin{equation} \hspace{-2.cm}
\begin{array}{l}    \phantom{  \sqrt{q_1\over q_2} K_{12} \cos{q_2y_2}  }  \Phi_1(x_1,y_1)   \cr   \phantom{  \sqrt{q_1\over q_2} K_{12} \cos{q_2y_2}  }  \Phi_2(x_2,y_3)  \cr
\phantom{  \sqrt{q_1\over q_2} K_{12} \cos{q_2y_2}    }
\Phi_3(x_3,y_3) \end{array}=
\begin{array}{lclrr}
u_1(x_1)   &\left[\sin{q_1y}\right. &+&                     K_{11}
\cos{q_1y_1}  &\left.\right]/(x_1y_1),  \cr u_2(x_2)   & &  &
\sqrt{q_1\over q_2} K_{12} \cos{q_2y_2}/(x_2y_2),  & \cr u_3(x_3)
& &  & \sqrt{q_1\over q_3} K_{13} \cos{q_3y_3}/(x_3y_3). & \cr
\end{array}
\end{equation}
We have used here the one-index notation and denoted by $q_i$  the
momentum of each channel.

More interestingly, and this was the main Faddeev's result, the T-matrix
perturbative expansion (\ref{T_pert}) can be reordered in such a way
that all delta functions disappear leading to compact equations.

\bigskip
To solve in practice equations (\ref{FE_Phi}) one
can perform a partial wave expansion of each FC in its proper set of Jacobi coordinates
\begin{equation}\label{PW_3}
  \Phi_i( \vec{x}_{i}, \vec{y}_{i})= \sum_{\alpha_i}
  {\varphi_{\alpha_i}(x_{i},y_{i})
  \over x_{i}y_{i}} {Y}_{\alpha_i}(\hat{x}_{i}, \hat{y}_{i}),
\end{equation}
%We use here the one-index notation for the Faddeev components, i.e:  $\Phi_i\equiv \Phi_{jk}$
where $Y_{\alpha_i}$ denote the bipolar spherical harmonics
\[{Y}_{\alpha_i}(\hat x_i,\hat y_i)=\sum_{m_xm_y}<l_xm_x;l_ym_y|l_xl_x;LM>Y_{l_xm_x}(\hat x_i)Y_{l_ym_y}(\hat y_i),\]
 and $\alpha_i=\{l_xl_yLM\}$ the set of intermediate quantum numbers in the angular (and eventually spin and isospin) couplings.
After inserting (\ref{PW_3})  in (\ref{FE_Phi}) and projecting the angular part, one is led with a system of two-dimensional coupled integrodifferential equations having the the form
\begin{eqnarray}
\left[ q^2+ \partial^2_{x}+ \partial^2_{y}  -v^e_{\alpha}(x,y)
\right] \varphi_{\alpha}(x,y)&=&v_{\alpha} (x) \cr &&\hspace{-5cm}
\sum_{\beta}  \int_{-1}^{+1} du\; h^{(3)}_{\alpha\beta}( x,y,u)
\varphi_{\beta}[x_{\beta}(x,y,u),y_{\beta}(x,y,u)],
\label{EFYPW_3}
  \end{eqnarray}
with  the effective potential
\begin{eqnarray*}
v^e_{\alpha}(x,y) &=& v_{\alpha}(x)  +{ l_{x_{\alpha}}(l_{x_{\alpha}}+1)\over x^2}
 +{ l_{y_{\alpha}}(l_{y_{\alpha}}+1)\over y^2}.
\end{eqnarray*}
The integral kernels $h^{(3)}_{\alpha\beta}$  constitutes the key
ingredient of the calculation. Their precise expressions can be
found in \cite{Gloeckle_Book_83,4B_h}.

Faddeev equations have a very simple form for 3 identical
particles in the so called S-wave approximation, i.e. where all
the angular momenta are set to zero:
\begin{equation}\label{S_wave}
    (q^2  + \partial^2_x + \partial^2_y -v_{\alpha})
    \phi_{\alpha}(x,y) = v_{\alpha} \sum_{\beta=1}^{n_a} c_{\alpha\beta}
    \int_{-1}^{+1} du \; {xy\over x'y' } \; \phi_{\beta}(x',y'),
      \end{equation}
with $\alpha,\beta=1,\ldots,n_a$,
\begin{eqnarray*}
2{x'}^2(x,y,u)&=& x^2 + 3y^2 -2\sqrt{3}xyu, \cr 2{y'}^2(x,y,u)&=&
3x^2 + y^2 +2\sqrt{3}xyu,
\end{eqnarray*}
and $c_{\alpha\beta}$ are numerical coefficients which take the values:
%\begin{itemize}
%\item
$n_a$=1 and $c$=2 for a 3  boson system with total angular momentum L=0,
%\item
 $n_a$=1 and  $c$=-1 for a 3  fermion system with total spin J=3/2,
%\item
$n_a$=2 and $c_{11}$=$c_{22}$=1/2, $c_{12}$=$c_{21}$=3/2 for a 3 fermion system  with total spin  J=1/2.
 %\(    \pmatrix{1/2 & 3/2 \cr 3/2 & 1/2} \)
%\end{itemize}

\bigskip
Before closing this section, several remarks are in order:

\begin{enumerate}
\item We have conscientiously ignored until now, the existence of
three-body terms  in the potential. They play an important role in
the description of nuclear systems but are not relevant in our
discussion. The interested reader can found a sound presentation in \cite{Gloeckle_Book_83}.

\item One of the essential properties of the Faddeev equations  in
view of fixing the boundary condition % puzzle  is due to
their decoupling in the asymptotic regions of the configuration space.
This is ensured under the assumption of short range pairwise
interactions. However in presence of Coulomb forces this
decoupling is not guaranteed, at least in  the above presented original form,
and the whole formalism could seem questionable. This problem was
solved by Merkuriev \cite{Merkuriev_80,Merkuriev_81}  by
introducing an artificial splitting of the Coulomb potential into a
short and long range  parts by means of a smooth cut-off function.
The long range part is kept in the left hand side of the Faddeev
equations to define the Coulomb asymptotes and the short range
parts appear in the right hand side ensuring the decoupling of the
equations. Merkuriev approach has proven to be very successful in
several nuclear physics problems like p-d reactions \cite{Papp} as
well as handling purely atomic problems
\cite{KCG_PRA46_1992,Valdes,Yakovl}.

\item In parallel with Faddeev work,  P.H. Noyes and collaborators
had independently proposed, and properly solved for the bound as
well as for elastic scattering cases, the differential form of the
same equations \cite{Noyes_Fideldey_1968}. They are sometimes denoted as Faddeev-Noyes
equations, specially in their simplified S-wave form
(\ref{S_wave}).

%%%%%%%%%%%%%%%%%%%
\item It is worth noticing that, even for the three-body case,
there remain many unresolved problems, for instance those related with the presence
of an infinite number of the open asymptotic channels.
In particular, the  very challenging problem of atomic anti-hydrogen production  slow antiproton
collisions with hydrogen atoms due to the presence of large number of the open asymptotic channels.
\end{enumerate}

The Faddeev solution of the quantum mechanical  3-body problem was
of paramount importance in theoretical physics and definitely set
the foundations for an {\it ab initio} solution of the many particle systems.

\begin{figure}[h!]
\begin{center}
\centering\includegraphics[width=3.cm]{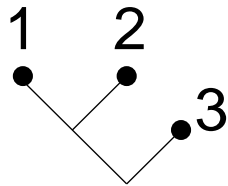}
\end{center}
\caption{Tree diagram representing the only possible way to build a 3-body cluster ending with an interacting pair (12)}\label{FY3_Tree}
\vspace{0cm}
\end{figure}

Nevertheless it remains a very particular case which somehow hides
the complexity of the  N-body  problem as well as the road for its
generalization to larger N. In the Faddeev equations, 3 is the
number of particles, the number of interacting pairs, the number
of Jacobi sets, as well as the number  of  ways for breaking (or
building!)  a 3 body cluster (123) into (or from) a lower rank
sub-clusters \cite{Faddeev_Birmingham_1969}. This later
characteristic,  of particular relevance in the N-body case,   can
indeed occur according  to the 3 different  "partitions"
(12)3,(13)2,(23)1 which have in fact the same topological
properties, i.e. that can be related to each other by permutation
operators. It is convenient to represent these different
topological types in form of "trees" diagrams, which in the N=3
case reduces to the single one represented in Fig. \ref{FY3_Tree}.

Thus, in this case, the number of particles encodes in fact very
different properties and make difficult to disentangle the
different role they play in the general theory. This is however
not the case in general as it will be illustrated in the following
sections.

%%%%%%%%%%%%%%%%%
\section{Faddeev-Yakubovsky equations for N=4}

Few years after Faddeev's pioneer work, Yakubovsky
\cite{Yakubovsky_SJNP5_1967}  proposed a consistent scheme to
build a set of equations which should allow the solution of the general N-body problem. His
demonstration was based on induction with respect to the number of
particles and this requires the previous solution of the N-1, N-2,
\ldots, 2 problems. After several unfruitful attempts
\cite{Weinberg_63_I_II}, Yakubovsky's result was a real tour de
force  to get rid of the $\delta$-like singularities in the
perturbative expansion of the N-body T-matrix, which disappear
after N-2 iterations of the proposed equations.

Rather than reproducing this general approach, which remain quite
abstruse even after some pedagogical efforts \cite{NY_1980}, we
will detail in this section --  without any formal demonstration
-- the main steps in deriving the differential equations for the N=4 case.

The starting point is  the intrinsic Schr\"{o}dinger's equation for
N=4:
\begin{equation}\label{Sch_4}
 (E-H_0)\Psi = V\Psi; \qquad  V= \sum_{i<j=1}^4 V_{ij} = V_{12}+V_{13}+V_{14}+V_{23}+V_{24}+V_{34},  \
\end{equation}
where the 4-body free Hamiltonian $H_0$ has  the
same spherical form (\ref{H0_3p_I}) as for N=3 case
\[ H_0 =   -  \left(   \Delta_{\vec{x}} +  \Delta_{\vec{y}} + \Delta_{\vec{z}}  \right),    \]
expressed in some set of Jacobi coordinates to be detailed later.

\begin{figure}[h!]
\begin{center}
\centering\includegraphics[width=6.cm] {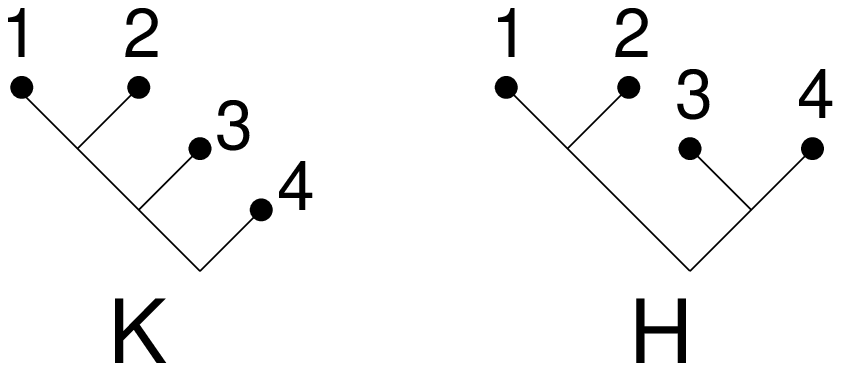}
\vspace{0.4cm}

\centering\includegraphics[width=6.cm]{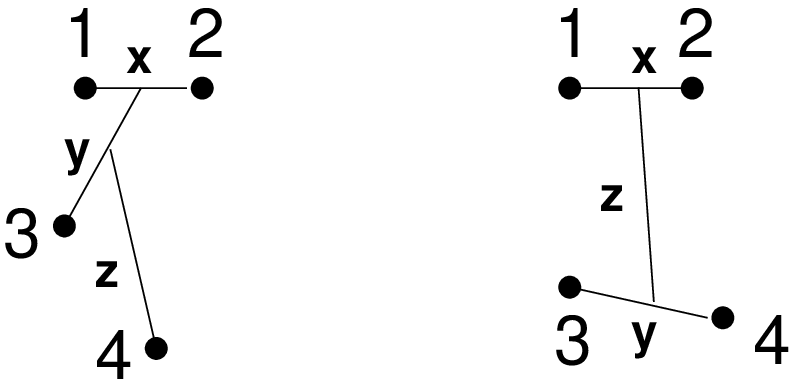}
\end{center}
\caption{Upper part: Tree diagram representing the two  possible topological ways for breaking fully interacting 4-body cluster
(1234) into substructures. Lower part: Jacobi coordinates corresponding to each cluster partition.} \label{FY4_H_Trees}
\vspace{0cm}
\end{figure}

The corresponding Faddeev equations (\ref{Phi_3p}) and (\ref{FE_Phi}) can be written in close analogy  with the N=3  case,   in terms of
the wave function partition
\begin{equation}\label{FC_4}
\Psi =  \sum_{i<j}\Phi_{ij}= \Phi_{12}+ \Phi_{13}+ \Phi_{14}+
\Phi_{23}+ \Phi_{24}+ \Phi_{34}.
\end{equation}
However, contrary to N=3, an additional decomposition of each
component ($\ref{FC_4}$) is needed to represent all different
asymptotes for the non interacting  particles. For instance
$\Phi_{12}$, associated with the  (12) interacting pair, and
fulfilling the Faddeev equation
\begin{equation} (E-H_0-V_{12} ) \Phi_{12}= V_{12}
 \left( \Phi_{13} + \Phi_{14}+ \Phi_{23}+ \Phi_{24}+ \Phi_{34}  \right),
  \end{equation}
is split into  3 components belonging  to two different types
\begin{equation} \Phi_{12} = \Phi_{12,3}^4 + \Phi_{12,4}^3 +
\Phi_{12,34},
\end{equation}
defined in terms of the interacting 2-body Green function
"embedded  in the 4-body space"
\[ G_{ij}=  ( E-H_0 -V_{ij})^{-1},   \]
by
\begin{eqnarray}
\Phi_{12,3}^4   &=&   G_{12} V_{12} ( \Phi_{13}+ \Phi_{23} ),\cr
\Phi_{12,4}^3   &=&   G_{12} V_{12} ( \Phi_{14}+ \Phi_{24} ),\cr
\Phi_{12,34}^4 &=&   G_{12} V_{12} \;\; \Phi_{34}.
\end{eqnarray}
They satisfy
\begin{eqnarray}
(E-H_0-V_{12})\Phi_{12,3}^4 &=&V_{12}\left( \Phi_{13}+\Phi_{23}
\right), \cr (E-H_0-V_{12})\Phi_{12,4}^3 &=&V_{12}\left(
\Phi_{14}+\Phi_{24}   \right),\cr (E-H_0-V_{12})\Phi_{12,34}
&=&V_{12}\Phi_{34}.
\end{eqnarray}
By repeating this procedure to each of the six Faddeev components
($\ref{FC_4}$)
\begin{equation}\label{FY_4}
 \Phi_{ij} = \Phi_{ij,k}^l + \Phi_{ij,l}^k + \Phi_{ij,kl},
 \end{equation}
one obtains the  set of 18 Faddeev-Yakubovsky (FY) components
defined by
\begin{eqnarray}
\Phi_{ij,k}^l &=& G_{ij} V_{ij} ( \Phi_{ik}+ \Phi_{jk}  ), \cr
%\Phi_{ij,l}^k &=& G_{ij} V_{ij} ( \Phi_{il}+ \Phi_{jl}  ) \cr
\Phi_{ij,kl}^l &=& G_{ij} V_{ik} \; \Phi_{kl},  \label{FYC_4}
\end{eqnarray}
and satisfying a system of 18 coupled equations
\begin{eqnarray}
(E-H_0-V_{ij})\Phi_{ij,k}^l&=&V_{ij}\left(
\Phi_{ik,j}^l+\Phi_{ik,l}^j+\Phi_{ik,lj} + \Phi_{jk,i}^l +
\Phi_{jk,l}^i + \Phi_{jk,il} \right), \cr
(E-H_0-V_{ij})\Phi_{ij,l}^k&=&V_{ij}\left(
\Phi_{il,j}^k+\Phi_{il,k}^j+\Phi_{il,kj} + \Phi_{jl,i}^k +
\Phi_{jl,k}^i + \Phi_{jl,ik}\right),\cr
(E-H_0-V_{ij})\Phi_{ij,kl}&=&V_{ij}\left(\Phi_{kl,i}^j+\Phi_{kl,j}^i+\Phi_{kl,ij}\right);
\hspace{1.cm}i<j,k<l.  \label{FY5_C}
\end{eqnarray}
These equation were first  formulated in momentum
space by Yakubovsky \cite{Yakubovsky_SJNP5_1967}, their
differential form with the corresponding boundary conditions was
elaborated some time later by Merkuriev and collaborators in
\cite{MY_1982,MY_1982_1983,MYG_NPA_1984}. Notice that in terms of
the total wave function the FY components read
\begin{eqnarray}
\Phi_{ij,k}^l &=& G_{ij} V_{ij} G_0( V_{ik}+ V_{jk}  ) \; \Psi,
\cr
%\Phi_{ij,l}^k &=& G_{ij} V_{ij} G_0( V_{il}+ V_{jl}  )  \;\Psi \cr
\Phi_{ij,kl}^l &=& G_{ij} V_{ik} G_0 \; V_{kl}    \; \Psi.
\end{eqnarray}
Each  FY component of $\Phi_{ij,k}^l$ or $ \Phi_{ij,kl}$
type\footnote{Usually denoted by $K_{ij,k}^l\equiv\Phi_{ij,k}^l$
and $H_{ij,kl}\equiv\Phi_{ij,kl}$} corresponds to a topologically
different way of breaking  a 4-body cluster into its subsystems up
to the point where a single interacting pair remains unbroken.
These sequences of breaking an N-body cluster -- denoted in the
literature "complete partition chains" -- are illustrated in Fig.
\ref{FY4_H_Trees} for the pair (12).

The upper left figure represents - from bottom to the top in
chronological order - a "tree" where a four particle system (1234)
by separating the particle 4 is first broken into 3+1 structure
[(123)4]. Then cluster (123) is decomposed by separating the
particle 3, thus building the 4-particle structure [(12)34]. This
corresponds to the K-like FY component $\Phi_{12,3}^4$ and to the
partition chain  (1234)$\supset$(123)4$\supset$(12)34.

In the upper right panel, an alternative partition strategy is demonstrated,
where  a four particle system (1234) is first broken by separating
pairs  (12) and (34), getting  [(12)(34)] partition. In the next
step the (34) pair is  decomposed [(12)34]. It corresponds to the
H-like FY component $\Phi_{12,34}$ and to the partition chain:
(1234)$\supset$(12)(34)$\supset$(12)34.

%The lower panels represent the  Jacobi coordinates associated to each partition.
 .

By permuting particle indexes, one obtains  6x2=12 K-like
components and 3x2=6 H-like, which give the 18 FY coupled
equations (\ref{FY5_C}). In case of identical particles
permutation symmetry allows to express all 12 K-like components
from any of them (idem for H-like components); thus one may reduce
the problem to two coupled equations for the two FYC  of fig Fig.
\ref{FY4_H_Trees}. The Jacobi coordinates, associated with each FY
component ($\vec{x}_K,\vec{y}_K,\vec{z}_K$) or
($\vec{x}_H,\vec{x}_H,\vec{x}_H$), are schematically represented
in the lower panel of Fig. \ref{FY4_H_Trees}. They are denoted
generically by $\vec{q}$ and have the general form given in
(\ref{Jacobi_3}), i.e.
\begin{equation}\label{Jacobi_G}
\vec{q}_{s,t} = \sqrt{ 2\mu_{st}\over m_0}  \left( \vec{R}_s-
\vec{R}_t  \right),
\end{equation}
where $\vec{R}_i$ denotes the center of mass positions of the two
clusters (eventually single particles) connected by each
coordinate and $\mu_{st}$ their reduced mass.

A practical solution of the FY equations  can be achieved  by
means of a partial wave expansion of the FY components in analogy
with the 3-body case (\ref{PW_3}). One ends with a system of
tridimensional integrodifferential equations of the form
\begin{eqnarray}
\left( q^2+   \Delta_{x,y,z} -v^e_{\alpha} \right) \varphi_{\alpha}(x,y,z)&=&
    v_{\alpha} \sum_{\beta}  \int_{-1}^{+1} du dv\;
    h^{(4)}_{\alpha\beta}( x,y,z,u,v),  \cr
&&\hspace{-4cm}  \varphi_{\beta}[x_{\beta}(x,y,z,u,v)
,y_{\beta}(x,y,z,u,v),z_{\beta}(x,y,z,u,v)],    \label{EFYPW_4}
\end{eqnarray}
with $\Delta_{x,y,z}=\partial^2_{x}+ \partial^2_{y} + \partial^2_{z} $ and the effective potential
\begin{eqnarray}
%\Delta_{x,y,z}&=& \partial^2_{x}+ \partial^2_{y} + \partial^2_{z} \cr
 v^e_{\alpha}& =& v_{\alpha}(x)  +{ l_{x_{\alpha}}(l_{x_{\alpha}}+1)\over x^2}
  +{ l_{y_{\alpha}}(l_{y_{\alpha}}+1)\over y^2}   +{ l_{z_{\alpha}}(l_{z_{\alpha}}+1)\over
  z^2},
  \end{eqnarray}
and $h^{(4)}_{\alpha\beta}$ some integral kernels which can be
found in \cite{4B_h}. For the case of 4 identical bosons in the
S-wave approximation former equations take the simple form
\cite{MYG_NPA_1984}
\begin{eqnarray}
( q^2+   \Delta_{x,y,z}-v) \varphi_1(x,y,z)&=&v(x)  \left[
\int_{-1}^{1}du    {xy \over x'y'_1}  \varphi_1(x',y'_1,z) \right.
\cr &+& \left. {1\over2}\sum_{i=1,2}\int\int_{-1}^{1}du dv
{xyz\over x'y''_iz''_i} \varphi_i(x',y''_i,z''_i)\right],\nonumber
\cr ( q^2+ \Delta_{x,y,z}-v)
\varphi_2(x,y,z)&=&v(x)\left[\varphi_2(y,x,z)  + \int_{-1}^{1}
dv\; {xz\over\hat{y}_1\hat{z}_1}
\varphi_1(y,\hat{y}_1,\hat{z}_1)\right],
\end{eqnarray}
where
\begin{eqnarray*}
x'^2       (x,y;u)     &=& {1\over4}x^2 + {3\over4}y^2 -
{1\over2}\sqrt{3}xyu, \cr {y'_1}^2   (x,y;u)     &=& {3\over4}x^2
+ {1\over4}y^2 + {1\over2}\sqrt{3}xyu,\cr {y''_1}^2
(x,y,z;u,v)&=&{1\over9}{y'_1}^2(x,y;u)+{8\over9}z^2 +
{4\over9}\sqrt{2}y'_1(x,y;u)zv,\cr {z''_1}^2
(x,y,z;u,v)&=&{8\over9}{y'_1}^2(x,y;u)+{1\over9}z^2 -
{4\over9}\sqrt{2}y'_1(x,y;u)zv,\cr {y''_2}^2
(x,y,z;u,v)&=&{1\over3}{y'_1}^2(x,y;u)+{2\over3}z^2 -
{2\over3}\sqrt{2}y'_1(x,y;u)zv,\cr {z''_2}^2
(x,y,z;u,v)&=&{2\over3}{y'_1}^2(x,y;u)+{1\over3}z^2 +
{2\over3}\sqrt{2}y'_1(x,y;u)zv,\cr \hat{y}_1^2(x,z;v)     &=&
{1\over3}x^2 + {2\over3}z^2 - {2\over3}\sqrt{2}xzv,   \cr
\hat{z}_1^2(x,z;v)     &=& {2\over3}x^2 + {1\over3}z^2 +
{2\over3}\sqrt{2}xzv.
\end{eqnarray*}

During the last twenty years these equations have been used to
solve very diverse bound state and scattering problems related
with nuclear or atomic physics,  including the break-up
reactions \cite{4B_Nogga,4B_Uzu,4B_Deltuva,4B_Laza}. The
interested reader can find in these references a detailed
description of the proper techniques to implement the boundary conditions and the associated numerical tools.

%%%%%%%%%%
\section{Faddeev-Yakubovsky equations for N=5}

As in  the previous cases, the solution of the corresponding
Schr\"{o}dinger equations starts with the Faddeev-like components
associated to each interacting pair
\begin{equation} \Psi
=\sum_{i<j} \Phi_{ij}=\Phi_{12}+
\Phi_{13}+\Phi_{14}+\Phi_{15}+\Phi_{23}+\Phi_{24}+\Phi_{25}+\Phi_{34}+\Phi_{35}
+\Phi_{45}, \end{equation} according to (\ref{Phi_3p}) and
(\ref{FE_Phi}). The number of interacting pairs, equal to the
number of Faddeev equations (\ref{FE_Phi}), is now 5x4/2=10.

\begin{figure}[h!]
\begin{center}
%\epsfxsize=8cm\centerline{\epsfbox{FY4_V.eps}}
%\centering\includegraphics[width=8.9cm]{FY4_V.eps}
\centering\includegraphics[width=11.5cm]{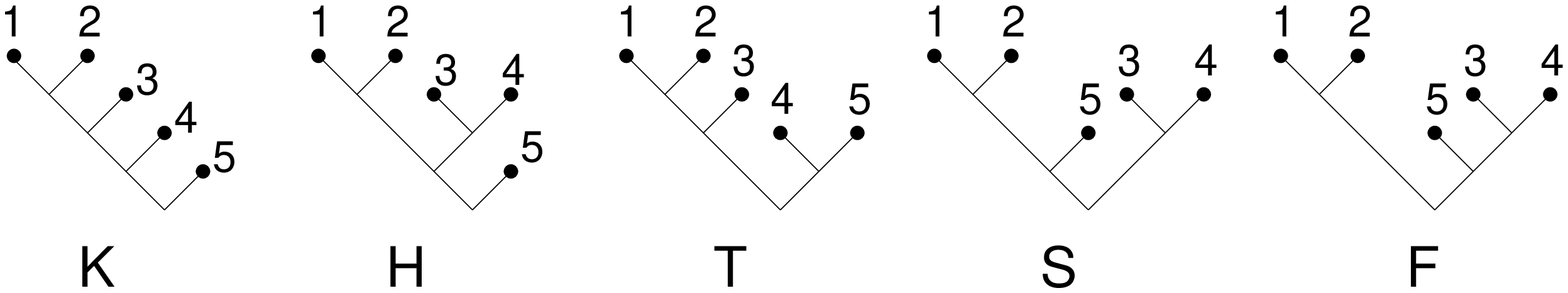}
\vspace{0.4cm}

\centering\includegraphics[width=11.5cm]{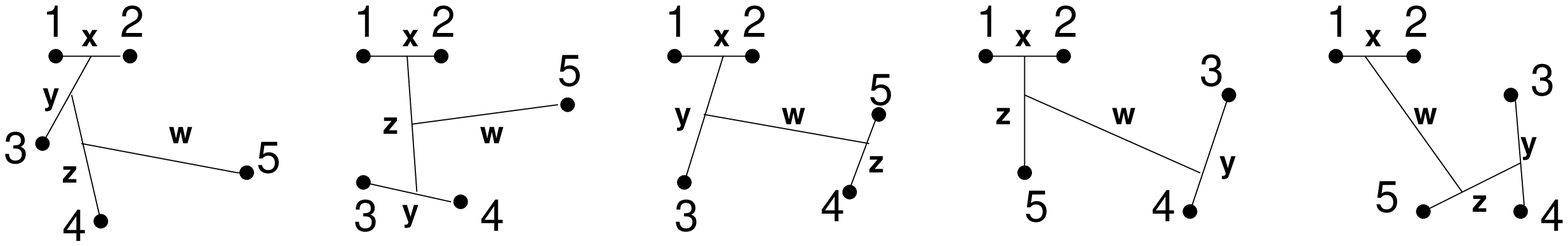}
\end{center}
\caption{The same as Fig  \ref{FY4_H_Trees} for the N=5 case.
Upper part: Tree diagram representing the five  possible
topological ways for breaking the fully interacting 5-body cluster
(12345) into substructures. Lower part: Jacobi coordinates corresponding to each cluster partition.}\label{FY5_H_Trees}
\vspace{0cm}
\end{figure}

Each of these ten  Faddeev components $\Psi_{ij}$ is decomposed
into six 4-body like Yakubovsky components
\begin{equation}\label{FY_4}
 \Phi_{ij} = \Phi_{ij}^{ijk} + \Phi_{ij}^{ijl}+\Phi_{ij}^{ijm} +        \Phi_{ij}^{ij,kl}   + \Phi_{ij}^{ij,km} +
 \Phi_{ij}^{ij,lm},
 \end{equation}
 defined, as in the N=4 case (\ref{FYC_4}), by
 \begin{eqnarray}
\Phi_{ij}^{ijk} &=& G_{ij} V_{ij} ( \Phi_{ik}+ \Phi_{jk}
),\nonumber \cr \Phi_{ij}^{ij,kl} &=& G_{ij} V_{ik} \; \Phi_{kl}.
\end{eqnarray}
In its turn, the  4-body components are further decomposed in
terms of five independent 5-body FY components -- denoted
following \cite{Rimas_5N_FBS_2018,Rimas_5N_PRC97_2018}  by
K,H,T,S,F -- in the following way:
\begin{equation}
\Phi_{ij} = \psi_{ij}^{ijk}  + \psi_{ij}^{ijl}  + \psi_{ij}^{ijm}
 + \psi_{ij}^{ij,kl}  + \psi_{ij}^{ij,km}    + \psi_{ij}^{ij,lm},
\end{equation}
with
\begin{eqnarray}
 \psi_{ij}^{ijk}      &=&   K_{ij,k}^l  + K_{ij,k}^m +   T_{ij,k},   \cr
  \psi_{ij}^{ij,kl}   &=&   H_{ij,kl}  + S_{ij,kl}  + F_{ij,kl}.   \label{psi_ijklm}
\end{eqnarray}
Each of these  5-body FY components corresponds  to a
topologically independent way of breaking a 5-body cluster. At
each step one of the "interacting" clusters is decomposed into two
pieces, giving rise to a corresponding FY equation. The procedure
is repeated until %the system is decomposed to the point where it
there remains
only a single "interacting" pair. The possible partition chains
for a 5-body system are represented in  the upper part of Fig.
\ref{FY5_H_Trees}. On the left side one may identify K-like and
H-like structures, which are associated with 4+1 particle
structures. The three components on the right hand side (T-like,
S-like and F-like) represent 3+2 particle structures.

The lower part of Fig. \ref{FY5_H_Trees}  represents the Jacobi coordinates associated to each FY component.
They have the general form (\ref{Jacobi_G}) and are denoted by
\[ \vec{q}= \{ \vec{x},\vec{y},\vec{z},\vec{w} \}. \]

By permuting the particle index one can see that there are 60
different K-type amplitudes, and 30 for each of the  other types:
H,T,S,F. Finally, this result into the set of 180 FY equations,
first derived in~\cite{Sasakawa_5b},
\begin{eqnarray}
(E-H_0-V_{ij} )  K_{ij,k}^l  &=& V_{ij}  \left[   K_{ik,j}^l   +
K_{jk,i}^l    +   \psi_{ik}^{ikl}   +   \psi_{jk}^{jkl}   +
\psi_{ik}^{ik,jl}  +   \psi_{jk}^{jk,jl}     \right],   \cr
(E-H_0-V_{ij} )  H_{ij,kl} &=& V_{ij}  \left[   H_{kl,ij}    +
\psi_{kl}^{kli}      +   \psi_{kl}^{klj}      \right],   \cr
(E-H_0-V_{ij} )  \;\;T_{ij,k} &=& V_{ij}  \left[   T_{ik,j}    +
T_{jk,i} +      \psi_{ik}^{ik,lm}    +      \psi_{jk}^{jk,lm}
\right],   \cr (E-H_0-V_{ij} )  S_{ij,lm} &=& V_{ij}  \left[
F_{lm,ij}     +      \psi_{lm}^{lm,jk}    +      \psi_{lm}^{lm,ik}
\right],   \cr (E-H_0-V_{ij} )  F_{ij,lm} &=& V_{ij}  \left[
S_{lm,ij}     +      \psi_{lm}^{lmk}    \right],
\end{eqnarray}
with  $\psi$'s given by (\ref{psi_ijklm}).

Each FY component $F_a$=(K,H,T,S,F) is expressed in its own Jacobi
set and take now the values on ${\mathbf R}^{12}$ and expanded in
spherical harmonics for each angular variable:
 \begin{equation}
F_a ( \vec{x},\vec{y },\vec{z},\vec{w})=   \sum_{\alpha}
{f_{a,\alpha}( x,y,z,w  )\over xyzw}  \;  \;  Y_{\alpha} (\hat{x},
\hat{y},\hat{z},\hat{w}),  \end{equation} where  $Y_{\alpha}$ is a
generalized "quadripolar harmonic" accounting for the angular
momentum, and eventually  spin and isospin, couplings
\begin{equation} Y_{\alpha} =     \left[   [ l_x, l_y ]_{l_{xy}}
\; [l_z,l_w]_{l_{zw}} \right]_{L},       \end{equation}
%\[  Y_{\alpha} =  \left[    \left[   [ l_x, l_y ]_{l_{xy}}  \; [l_z,l_w]_{l_{zw}} \right]_L  \; \{S\}_S    \right]_{J} \{T\}_{TT_z} ,  \]
%\[ \{S \}_S\!=\mid \![[s_1,s_2]_{s_x}  [s_3,s_4]_{s_y}]_{s_{xy}} s_5 ;SS_z\rangle\]
%\[  \{T \}_{TT_z} \!= \mid \! [[t_1,t_2]_{t_x}  [t_3,t_4]_{t_y}]_{t_{xy}} t_5 ;TT_z\rangle\]
and $\alpha$ labels the set of quantum numbers involved in the intermediate couplings.

After projecting the angular part, one is left with a set of
coupled four-dimensional integro-differential equations for the
reduced radial amplitudes $f_{a,\alpha}$ of the form
\begin{eqnarray}
 \left[  q^2 + \Delta_{xyzw} - v^e_{\alpha}(x)  \right]  f_{a,\alpha}(x,y,z,w)=  v_{\alpha}(x)\sum_{b\beta} && \hspace{-0cm}\cr
\int\int\int d\theta d\xi d\zeta \;
h^{(5)}_{a\alpha,b\beta}(x,y,z,w,\theta,\xi,\zeta)
f_{b,\beta}(x_b,y_b,z_b,w_b),&&  \label{EFYPW_5}
\end{eqnarray}
 with $\Delta_{xyzw} = \partial^2_x+ \partial^2_y+\partial^2_z+\partial^2_w $ and the effective potential
\[v^e_{\alpha}(x) =v_{\alpha}(x)  - {l_x(l_x+1)\over x^2}-  {l_y(l_y+1)\over y^2} - {l_z(l_z+1)\over z^2} - {l_w(l_w+1)\over w^2}. \]

These equations have been solved for the first time in a recent
work devoted to study the n-$^4$He scattering
\cite{Rimas_5N_PRC97_2018}, more recently they were applied to
compute the complex energies of the $^5H$  resonant states
\cite{LHC_PLB791_2019}

%%%%%%%%%%%%%%%%%%%%%%%%%%%
\section{Numerical methods}

In the last years we have developed a numerical protocol to solve
the FY equations in configuration space for N=3,4 and 5 particles.
For the sake of simplicity we will illustrate it in the case N=3;
the close analogy we kept in the final equations
 (\ref{EFYPW_3}), (\ref{EFYPW_4}) and (\ref{EFYPW_5}), make  the
 generalization rather straightforward.

In N=3, we wish  to determine on a two dimensional
domain $D=[0,x_{n_x}]\times[0,y_{n_y}]$ the partial wave amplitudes
of the FY components which are solution of
eq. (\ref{EFYPW_3}).
%\begin{eqnarray}
%\left[   q^2 + \Delta_{xy}  -v^e_{\alpha}(x) \right] \varphi_{\alpha}(x,y) &=&v_{\alpha}(x) \cr
%\sum_{\alpha'} \int_{-1}^{+1} du h_{\alpha\alpha'} (x,y,u)     \varphi_{\alpha'}[x'_{\alpha'}(x,y,u),y'_{\alpha'}(x,y,u)] $$
%\end{eqnarray}
Our unique approximation is the assumption that the solution we are
looking for can be locally expanded in terms of some polynomial basis:
\begin{equation}\label{Spline}
 \varphi_{\alpha}(x,y)= \sum_{i=0}^{N_x} \sum_{j=0}^{N_y}
  c_{\alpha, ij} f_{i} (x)f_{j}(y).
\end{equation}
We used two kind of such bases: the so called splines
$f_i(x)\equiv S_i(x)$ (cubic or quintic)
 and the Lagrange interpolating functions $f_i(x)\equiv L_i(x)$.
In what follows we will particularise the case of cubic splines.
The interested reader can find a detailed explanation of both
choices in Sect. 2.9 of  Ref. \cite{Rimas_DHDR_2019}.

The cubic spline functions $S_i(q)$ are associated to each
variable $q=x,y$, they are constructed upon a ($n_q$+1)-point grid
$G_q=\{q_0,q_1,\ldots,\ldots q_n\}$ defined on each of the
intervals $[q_0=0,q_n]$ of the resolution domain  $D$.

Two  splines are associated to each grid point $q_i$:  $S_{2i}$
and $S_{2i+1}$. Both have  a finite support in the two consecutive
intervals $[q_{i-1},q_i]\cup[q_{i},q_{i+1}]$, are piecewise cubic
polynomials on each of them and have $C^1$ matching between them.
They have the useful properties:
\[ S_{i}(q_j)= \delta_{i,2j};  \qquad  S'_{i}(q_j)= \delta_{i,2j+1};   \qquad \forall j=0,n_q.\]

By inserting (\ref{Spline}) into eq. (\ref{EFYPW_3}) and
validating on an ensemble of $N_x\times N_y\equiv(2n_x+2)\times
(2n_y+2)$ well chosen points $\{ \bar{x}_i \bar{y}_j\}$ for each
FC one obtains a linear system allowing to determine the unknown
coefficients of the expansion (\ref{Spline}):
\begin{equation}\label{LcRc}
Lc=Rc \quad \Longleftrightarrow \quad
 \sum_{\alpha' i' j' } L_{\alpha ij , \alpha' i' j' } \;
 c_{ \alpha' i' j' }   = \sum_{\alpha'  i' j' } R_{\alpha ij ,
  \alpha' i' j' } \; c_{ \alpha' i' j' }.
\end{equation}
with  $i,i'=0,\ldots 2n_x+1$ and $j'=0,\ldots 2n_y+1$ and
$\alpha,\alpha'=1,\ldots n_a$ where $n_a$ is the number of partial
wave amplitudes of the Faddeev component $\phi_a$. Usually, by
considering the fact that Faddeev components are regular at the
origin, the $i=0$ and $j=0$ splines might be neglected as they
will require coefficients with null values. Then the dimension of
the linear system for N=3 is $d=3 \times n_a \times
(2n_x+1)(2n_y+1)$.

The matrix elements of the  left hand side are generalized by
\begin{eqnarray*}
L_{\alpha ij,\alpha'i'j'}
=\delta_{\alpha\alpha'}\phantom{BBBBBBBBBBBBBBBB} &&\cr \left[ q^2
S_{i'} (\bar x_i)S_{j'}(\bar y_j)  + S^"_{i'} (\bar
x_i)S_{j'}(\bar y_j) + S_{i'} (\bar x_i)S^"_{j'}(\bar y_j) -
v^e_{\alpha}(\bar x_i) S_{i'} (\bar x_i)S_{j'}(\bar y_i) \right],
\end{eqnarray*}
and those of the right hand side by
\[  R_{\alpha ij,\alpha'i'j'} =   v_{\alpha}(\bar x_i)  \int_{-1}^{+1} du\; h_{\alpha\alpha'} (\bar x_i,\bar y_j, u)    \; S_{i'} [x'_{\alpha'}(\bar x_i ,\bar y_j ,u)  ] \; S_{j'}[y'_{\alpha'}(\bar x_i ,\bar y_j ,u) ]. \]

In the scattering problems, the boundary conditions introduce an
inhomogeneous term in the right hand side of (\ref{LcRc}) and one
is left with solving a linear system,  generalized as
\begin{equation}
 Ax=b; \qquad A=L-R, \end{equation}
where $x$ holds for the
unknown coefficients $c_{\alpha ij} $. However the bound state
problem is an homogeneous one
\begin{equation} A x = \lambda x,\end{equation}
where both $\lambda$ and $x$ are unknowns. Both problems can be
unified by using the inverse iteration method. It consist in
solving iteratively the inhomogeneous system
\[ (A-\lambda_0)x^{(k+1)}=x^{(k)}; \quad k=0,1,....nite. \]
starting with a trial value $\lambda_0$ and an initial guess $x^0$.
One can show that under some conditions the series
 $x^{(0)} , x^{(1)}, x^{(2)},\ldots $ converges
towards the  eigenvector of $A$ whose eigenvalue
is closest to the trial value $\lambda_0$.
By doing so we are always reduced to solving an
 inhomogeneous linear system.

In view of solving very large linear systems
\begin{equation}\label{Axb}
Ax=b.
\end{equation}
the direct methods (Gauss elimination, LU decomposition,...) are
not applicable for they require storage capabilities beyond the
present technology and turns to be very slow. We use alternative
methods, based also on iterative procedures which, starting from
an "educated guess" $x^{(0)}$, generate a series $x^{(i)}$ which
minimize the residual $r^{(i)}=\mid Ax^{(i)}-b\mid$ until we
consider it to be "small enough". They are based on the
matrix-vector multiplication operations which require only to
store a small amount of data, which allows reconstruction of the
matrix A elements on the fly.

There are several families of iterative algorithms \cite{Saad_2003}.
The so called Bi-Conjugate Gradient Stabilized (BICGSTAB)
 is a very robust one that we have extensively used.

One of the limitation of the iterative methods is however the
number of iteration required until convergence. Even in the 3-body
problem with $d\sim 10^5$ this number is prohibitive. To overcome
this difficulty one uses the preconditioning technique,
 which consist in
finding an approximation of the inverse matrix $A^{-1}$, say $\hat{A}^{-1}$,
and solve instead of
(\ref{Axb}) the equivalent system
\begin{equation}  \hat{A}^{-1} Ax= \hat{A}^{-1}b. \end{equation}
We have systematically used this technique taking as a
preconditioning
 matrix  the one appearing in the left-hand-side of equations, the matrix  $\hat{A}=L $.
Its inversion can be performed exactly by means of the so called
 "tensor trick", introduced by \cite{Kok}.
It is based on the fact that the matrix $L$ is a sum of 4 terms
having a tensorial structure \(L_i= P_i \otimes  Q_i \). We will
illustrate this procedure in the case of one single amplitude
\begin{eqnarray*}
 L_{ij,'i'j'} &=&  q^2 S_{i'} (\bar x_i)S_{j'}(\bar y_j)  - v(\bar x_i) S_{i'} (\bar x_i)S_{j'}(\bar y_j)     \cr
&+& S^"_{i'} (\bar x_i)S_{j'}(\bar y_j)  -  {l_x(l_x+1)\over \bar
x_i^2} \;   S_{i'} (\bar x_i)  S_{j'}(\bar y_j)  \cr &+& S_{i'}
(\bar x_i)S^"_{j'}(\bar y_j)  -  {l_y(l_y+1)\over \bar y_j^2} \;
S_{i'} (\bar x_i)  S_{j'}(\bar y_j),
 \end{eqnarray*}
One easily identifies the following tensorial structure
\[ L=  L^x \otimes N^y + N^x \otimes L^y, \]
where the factors
\begin{eqnarray*}
N^x_{ii'} &=& S_{i'} (\bar x_i),
 \cr N^y_{ii'} &=&S_{j'}(\bar y_j),
\cr L^x_{ii'} &=&  q^2 S_{i'} (\bar x_i)  + S^"_{i'} (\bar x_i) -
{l_x(l_x+1)\over \bar x_i^2} \;   S_{i'} (\bar x_i)  -  v(\bar
x_i) S_{i'} (\bar x_i), \cr L^y_{jj'} &=&  S^"_{j'}(\bar y_j)  -
{l_y(l_y+1)\over \bar y_j^2} \;   S_{j'}(\bar y_j),
\end{eqnarray*}
have dimensions usually equivalent to  the square root of the
original matrix $L$. Let us rewrite $L$ it in the form
\begin{eqnarray*}
L = N_x\otimes N_y \cdot ( N_x^{-1}\cdot L_x\otimes{\bf 1_y} +
{\bf 1_x}\otimes N_y^{-1}\cdot L_y ),
\end{eqnarray*}
and diagonalize in $\mathbf C$
\begin{eqnarray*}
N_x^{-1}\cdot L_x &=&   U_x \cdot D_x \cdot U_x^{-1},      \cr
N_y^{-1}\cdot L_y &=&   U_y \cdot D_y \cdot U_y^{-1},  \cr
\end{eqnarray*}
with U unitary and D's diagonal matrices.
We obtain
\begin{eqnarray*}
L = N_x\otimes N_y \cdot ( U_x \cdot D_x \cdot U_x^{-1}
\otimes{\bf 1_y} + {\bf 1_x}\otimes    U_y \cdot D_y \cdot
U_y^{-1} ),
\end{eqnarray*}
which can be written in the form
\begin{equation}
 L=N_x\otimes N_y \cdot U_x\otimes U_y\cdot  (D_x\otimes{\bf 1_y}+{\bf 1_x}\otimes D_y)\cdot U_x^{-1}\otimes
 U_y^{-1},
\end{equation}
The inverse matrix is then easily obtained
\begin{equation}
 L^{-1}=  U_x\otimes U_y \cdot (D_x \otimes{\bf 1_y}+{\bf 1_x}\otimes D_y)^{-1} \cdot   U_x^{-1}\otimes U_y^{-1}  \cdot N_x^{-1} \otimes
 N_y^{-1},
\end{equation}
with
\[ (D_x\otimes{\bf 1_y}+{\bf 1_x}\otimes D_y)^{-1}_{ij,i'j'} = ( D^x_{i}+D^y_{j})^{-1}\delta_{ii'}\delta_{jj'}. \]

This procedure, that can be generalized to N=4
\cite{Rimas_PhD_2003} and N=5 case, allows us to obtain a
satisfactory preconditioning of the linear system (\ref{Axb}) and
a final solution  with a reasonably small number of iterations
($\sim 20$).

The numerical protocol  presented above involves as unique
assumption the local expansion of the solution (\ref{Spline}) in
terms of cubic (eventually quintic) polynomials. Its accuracy relies
only on the number of grid points ($n_x,n_y,\ldots$) for each
coordinate $q=x,y,\ldots$ as well as in the number of partial wave
amplitudes ($n_a$) on which the Faddeev-Yakubovsky
components are expanded. The convergence of the results is studied by
systematically increasing  these numbers.

\section{Summary}

We have  attempted to describe, in a  unified scheme of increasing complexity,
the Faddeev-Yakubovsky equations for the N=3, 4 and 5-body problems in configuration
space, as well as the numerical methods allowing its solution.

They are based on a recursive splitting of the intrinsic N-body
wave function, solution of the Schr\"{o}dinger
equation, in to as many components as there exist independent ways
(or "complete partition chains") to break fully connected
N-particle system into disconnected ones with a single interacting
particle pair remaining. They are represented by the "tree"
diagrams of Figs. \ref{FY3_Tree}, \ref{FY4_H_Trees} and
\ref{FY5_H_Trees}. To each diagram is associated a function, named
Faddeev-Yakubovsky component,  which takes the values on $\mathbf
R^{3(N-1)}$. The pioneering work of L.D. Faddeev for N=3
\cite{Fad_JETP39_1960} and its extension to an arbitrary N
achieved by O.A. Yakubovsky \cite{Yakubovsky_SJNP5_1967}, both
from the  Leningrad/Saint Petersburg University, formulated the
equations fulfilled by the ensemble of these components.

The original works of L.D. Faddeev and O.A. Yakubovsky  were
formulated  in momentum space, in terms of T-matrix partitions,
and solved under this form by Gl\"{o}ckle and collaborators for N=3 and N=4 \cite{Gloeckle_et_al}
The formulation in configuration space was due to
 S.P. Merkuriev and S. Yakovlev in a fruitful
 collaboration with the Grenoble theory group
of the former Institut des Sciences Nucl\'{e}aires
\cite{Merkuriev_71,Merkuriev_74,GLM_PRL33_1974,MGL_AP99_1976,MY_1982,MY_1982_1983,MYG_NPA_1984}.

In configuration space,  the Faddeev-Yakubovsky equations
result into a system of partial derivative equations coupling the
ensemble of the Faddeev-Yakubovsky components. After the partial wave
expansion is performed, they turn into a coupled system of
integro-differential equations on $\mathbf R^{N-1}$ with some
smooth (N-2)-dimensional integral kernels which can be solved by
standard linear algebra methods. Their generalization is  natural,
as it can be seen from (\ref{EFYPW_3}), (\ref{EFYPW_4}) and
(\ref{EFYPW_5})

The Faddeev-Yakubovsky equations  provide a mathematically
rigorous approach for the full solution of the N-body problem.
However its scalability with the number of interacting particles
is a serious drawback that dramatically  limits its range of
applicability. The fact that the formal object to study is a
function with arguments on ${\mathbf R}^{3(N-1)}$ is a first sign
of it, but not the only one. The situation, discussed in some
detail in \cite{Rimas_5N_FBS_2018}, is well illustrated in Table
\ref{Table_N} where the dimension of the N-body solution is
detailed in terms of number of equations, partial wave amplitudes,
and the linear algebra problem.

We would like to notice that other rigorous schemes have been
proposed for the solution of the N-body problem. One of them is
the  AGS equations \cite{AGS_1967} which have provided accurate
results for the 3- and 4-nucleon problem \cite{Fonseca_Deltuva}.

It is worth emphasizing also that such a rigorous mathematical
schemes are not necessary when dealing with bound states or simple
$1+(N-1)$ elastic scattering processes. The Schr\"{o}dinger
equation can then be directly solved by several methods, like GFMC
\cite{GFMC}, with Hyperspherical Harmonics \cite{Mario,Pisa} or
Gaussian basis  \cite{Emiko}, NCSM \cite{NCSM,dt_Nature_2019},
Lorentz Integral Transform \cite{LIT_94} which produces also very
accurate results, in some cases well beyond the technical
capabilities of the Faddev-Yakubovsky approach (see
\cite{Orlandini_2017} for a more detailed review). However, the
Faddeev-Yakubovsky partition of the wave function  is
interesting to increases the  numerical convergence of the results
or even unavoidable for an appropriate implementation of the
boundary conditions \cite{Emiko,Pisa_Scat}.

\begin{table}[h!]
\caption{Scalability of the Faddeev-Yakubovsky scheme as a
function of the number of interacting particles N.: ne is the
number of equations,(equal to the number of FY components), neid
the same number in case of identical particles, $n_a$ the number
of
amplitudes  in the partial wave expansion.}\label{Table_N}       % Give a unique label
\begin{tabular}{lr r  r }
\hline\noalign{\smallskip}
N& Ne                                      &  Neid                        & $n_a$  \\
\noalign{\smallskip}\hline\noalign{\smallskip}
2 & 1       & 1      &  2         \\
3 & 3       & 1      &  $\sim10^2$    \\
4 & 18      & 2      &  $\sim10^4$    \\
5 & 180     & 5      &  $\sim10^6$    \\
6 & 2700    & 15     &            \\
  &  \ldots &        &            \\
N & ${N! (N-1)! \over 2^{N-1}}$& Int$\left({2(N-1)!\over(\pi/2)^N}\right)$  & \\
\noalign{\smallskip}\hline
\end{tabular}
\end{table}

%\begin{acknowledgements}
%\end{acknowledgements}



\end{document}